\documentclass[aps,prb,superscriptaddress,amsmath,amssymb,floatfix,twocolumn]{revtex4-1}
\usepackage{times}
\usepackage{graphicx}
\usepackage{subfigure}
\usepackage{color}
\usepackage{epstopdf}

\begin{document}

\title{Orbital-selective Mott phase in multiorbital models for iron pnictides and chalcogenides}

\author{Rong Yu}
\affiliation{Department of Physics, Renmin University of China, Beijing 100872, China}
\affiliation{Department of Physics and Astronomy, Shanghai Jiao Tong
University, Shanghai 200240, China and Collaborative Innovation
Center of Advanced Microstructures, Nanjing 210093, China}
\author{Qimiao Si}
\affiliation{Department of Physics \& Astronomy, Rice University, Houston, Texas 77005}

\begin{abstract}
There is increasing recognition that the multiorbital nature of the $3d$ electrons is important to  the proper description of the electronic states
in the normal state of the iron-based superconductors.  Earlier studies of the
pertinent multiorbital Hubbard models
identified an orbital-selective Mott phase, which anchors the orbital-selective behavior
seen in the overall phase diagram. An important characteristics of the models is that the orbitals
are kinetically coupled -- {\it i.e.} hybridized -- to each other, which makes the orbital-selective Mott phase
especially nontrivial.
A $U(1)$ slave-spin method was used to analyze the model with nonzero orbital-level splittings.
Here we develop a Landau free-energy functional to
shed further light on this issue.
We put the microscopic analysis from the $U(1)$ slave-spin approach in this perspective,
and show that the intersite spin correlations are crucial to the renormalization
of the bare hybridization amplitude towards zero
and the concomitant realization of the orbital-selective Mott transition.
Based on this insight, we discuss additional ways to study the orbital-selective Mott physics from
a dynamical competition between the interorbital hybridization and collective spin correlations.
Our results demonstrate the robustness of the orbital-selective Mott phase in the multiorbital models
appropriate for the iron-based superconductors.
\end{abstract}

\maketitle


\section{Introduction}
In many strongly correlated systems, superconductivity is closely connected to a strongly-correlated bad-metal normal state
and a nearby antiferromagnetic order. As such, there has been considerable effort devoted to the understanding
of the electron correlation effects and the associated magnetism. In the case of the iron-based superconductors (FeSCs)
~\cite{Kamihara2008,Johnston,Dai2015,Si-NRM2016, Hirschfeld2016, FWang-science2011},
recent developments have further highlighted the importance of electron correlations.
For instance, angle-resolved photoemission spectroscopy (ARPES)
has
found an orbital-selective Mott phase (OSMP) in the iron chalcogenides
~\cite{MYi.2013,MYi.2015,YJPu.2016,MYi.2016}.
The OSMP phase arises in multiorbital Hubbard models of the FeSCs,
which contain both Hubbard interactions and Hund's coupling \cite{Yu2013,Yu-Cossms2013}.
Evidence for the OSMP has also come from a variety of other measurements~\cite{Wang2014,Ding2014,Li2014,Gao2014}.
A complementary approach to the electron correlations of the FeSCs
describes the localization-delocalization phenomena in the form of an orbital differentiation
and a coherence-incoherence crossover~\cite{Yin2011,Aron2015},
though OSMP is not explicitly invoked.
In addition,
recent experiments have identified a Mott insulating system in the copper-doped iron pnictides \cite{Song2016},
which accompanies the earlier observations of Mott insulating states in the iron chalcogenides \cite{Fang2011,Zhu2010,Birgeneau2015}.
These observations considerably expand on the bad-metal behavior known,
since early on in the field, through
other properties of the iron pnictides \cite{Basov.2009,Si2008,Haule08}.
For example, the room-temperature electrical resistivity is large (reaching the Mott-Ioffe-Regel limit, as defined by the normalized mean free path
$k_F \ell$ being of order unity), in contrast to good metals such as Cr (for which $k_F \ell$ at  room temperature is much larger than $1$).
Moreover, the Drude optical weight is much reduced\cite{Basov.2009}.

The parent systems of FeSCs have the iron valence $+2$, with $n=6$ electrons occupying the 3$d$ orbitals.
Correspondingly, multiple 3$d$-electron orbitals are important~\cite{Graser09,Si2008,Si-NJP09,Kuroki08,Cao08,Haule08,LeeWen08}.
In order to address the bad-metal physics, an important question is how the itinerancy of the electronic states is reduced with increasing
Coulomb interactions.  The appropriate multiorbital Hubbard models
include the intra- and interorbital Hubbard interactions ($U$ and $U'$) and the Hund's coupling ($J_{\rm{H}}$).

The OSMP phase appears in the paramagnetic solution to the multiorbital Hubbard models \cite{Yu2013}.
In this phase,
the non-degenerate $3d$ $xy$ orbital loses its coherent spectral weight at the Fermi energy,
while this weight remains nonzero for the other orbitals (including the degenerate $3d$ $xz/yz$ orbitals).
These features have been clearly identified in the ARPES measurements \cite{MYi.2013,MYi.2015,YJPu.2016,MYi.2016}.

From a theoretical perspective, it is important to stress that,  the orbitals are kinetically coupled
-- {\it i.e.} hybridized -- to each other in these models.
In particular, the bare kinetic hybridization between the $3d$ $xy$ orbital and the other $3d$ orbitals is nonzero.
In the OSMP phase, this hybridization is renormalized to zero and the $3d$ $xy$ orbital is no longer mixed
with the other orbitals in the low-energy electronic excitations.

More generally, orbital selectivity has been discussed
in the correlation effects; see, for example, Refs.~\onlinecite{Yu2011,Yin2011,Yu-U12012, Bascones2012,deMedici2014,Zhang2012}.
It can be defined in terms of the mass enhancement ($m^*/m_b$, the ratio of the effective mass observed experimentally to that
of the non-interacting band dispersion) being different for the
electronic states with predominantly different orbital contents.
In practice, $m^*/m_{b}$ for electronic states with predominantly $3d$ $xy$ orbital character is much larger (reaching $10-20$
in the iron selenides) than that for the electronic states with predominantly $3d$ $xz/yz$
(typically $3-4$) and other orbital characters \cite{Si-NRM2016,MYi.2016}.

Strictly speaking, the orbital selectivity is in itself not precisely defined,
because the hybridization mixes the different orbitals in the electronic band states.
However, it becomes sharply defined as a precursor to the OSMP \cite{Yu2013}.
In other words, the OSMP anchors the notion of orbital selectivity in general.
Note that the picture of the orbitally-differentiated coherence-incoherence crossover
has also yielded the $3d$ $xy$ orbital to be localized at sufficiently high temperatures,
while itinerant at zero temperature
~\cite{Yin2012,Miao2016}.
Hence the generalization of the notion of OSMP is relevant in this case as well.
These highlight
 the conceptual importance of the OSMP in characterizing the orbital selectivity in general.

\subsection{Orbital vs. band basis}

OSMP for the multiorbital Hubbard model has been discussed in several contexts.
Most of these studies start from multiple inequivalent bands,
and define the Hubbard and Hund's interactions in the band states \cite{Anisimov2002,deMedici2009,deMedici2011}.
By definition, the band basis diagonalizes the noninteracting part of the Hamiltonian. In other words, kinetically,
they are decoupled from each other. In the presence of interactions, the weight of the
coherent electrons near the Fermi surface will be renormalized below $1$ in a
band-dependent way. An OSMP corresponds to the regime where the renormalized coherent spectral weights for some
of the bands vanish while those for the others remain nonzero.

For the FeSCs, multiorbital Hubbard models are defined in the orbital basis
~\cite{Yu-Cossms2013, Kuroki08, Si2008, Si-NJP09, Cao08, Graser09, Haule08,LeeWen08,Yu2011,Yu-U12012, Bascones2012,deMedici2014,Zhang2012}.
The definition of the non-degenerate orbitals is unambiguous.
For the degenerate $3d$ $xz/yz$ orbitals, there can be alternative definitions,
but the degeneracy ensures that
the different definitions are equivalent to each other.

In the orbital basis, the kinetic part of the Hamiltonian is not diagonal. Therefore,
in the Hamiltonian,
the orbitals are kinetically coupled -- {\it i.e.} hybridized -- to each other.
In the OSMP solution of Ref.~\onlinecite{Yu2013},
the destruction of the
quasiparticle spectral weight, $Z_{xy}$,
is accompanied by the suppression
of the renormalized interorbital kinetic hybridization
 between the $3d$ $xy$ orbital and the other ones.

The purpose of this paper is to clarify how the above happens. We do so by formulating a Landau free-energy functional, which also
demonstrates the robustness of the OSMP. Viewed from this perspective,
we show the crucial role that the intersite spin correlations
play in generating the OSMP ({\it cf.} Fig.~\ref{fig:1}) within the  the microscopic $U(1)$ slave-spin approach~\cite{Yu-U12012,Yu2013}.

The remainder of the paper is organized as follows. In Sec.~\ref{Sec:Model}, we define the model and summarize the $U(1)$ slave-spin approach.
Sec.~\ref{Sec:Landau} is devoted to the formulation of the Landau free-energy functional in terms of the orbital-dependent quasiparticle weight
and how the OSMP appears as a distinct phase permitted by the Landau functional.
We then address, in Sec.~\ref{Sec:OSMP-U1SS}, the OSMP as derived by
 the microscopic $U(1)$ slave-spin approach
from the perspective of the Landau free-energy functional.
Landau analysis has been useful in clarifying the Mott transition in the dynamical mean-field theory (DMFT) context~\cite{Kotliar1999}.
Sec.~\ref{Sec:Discussions} discusses the implications of the insight gained in the present work for further studies on the OSMP,
and Sec.~\ref{Sec:Conclusion} summarizes the paper. Finally, in Appendices~\ref{Sec:Appen_A} through \ref{Sec:Appen_C},
we present further details on the saddle-point equations
of the $U(1)$ slave-spin approach.

\section{Multiorbital Hubbard model and the $U(1)$ slave-spin approach}
\label{Sec:Model}

We now define the model and, to facilitate the analysis in the next section, summarize the microscopic approach~\cite{Yu-U12012}
based on a
$U(1)$ slave-spin representation.

\subsection{Multiorbital Hubbard model}

The multiorbital Hubbard model for the FeSCs takes the following form,
\begin{equation}
 \label{Eq:Ham_tot}
 H=H_0 + H_{\mathrm{int}}.
\end{equation}
$H_0$ contains the tight-binding parameters among the multiple orbitals,
\begin{equation}
 \label{Eq:Ham_0} H_0=\frac{1}{2}\sum_{ij\alpha\beta\sigma} t^{\alpha\beta}_{ij}
 d^\dagger_{i\alpha\sigma} d_{j\beta\sigma} + \sum_{i\alpha\sigma} (\Delta_\alpha-\mu) d^\dagger_{i\alpha\sigma} d_{i\alpha\sigma},
\end{equation}
where $d^\dagger_{i\alpha\sigma}$ creates an electron in orbital $\alpha$ ($=1,...,5$)
with spin $\sigma$ at site $i$, $\Delta_\alpha$
refers to the energy level associated with the crystal field splitting (which is diagonal in the orbital basis),
and $\mu$ is the chemical potential.
In general, $t^{\alpha\beta}_{ij}\neq0$ for $\alpha\neq\beta$, corresponding to a nonzero kinetic hybridization between
the different orbitals.
For latter references, we note that the onsite energy for the $3d$ $xy$ orbital is different from any
of the other four $3d$ orbitals: for any orbital $\beta \ne xy$, the level splitting $\Delta_{xy,\beta}
\equiv \Delta_{xy} - \Delta_{\beta} \ne 0$.
The onsite interaction $H_{\rm{int}}$ reads
\begin{eqnarray}
 \label{Eq:Ham_int} H_{\rm{int}} &=& \frac{U}{2} \sum_{i,\alpha,\sigma}n_{i\alpha\sigma}n_{i\alpha\bar{\sigma}}\nonumber\\
 &&+\sum_{i,\alpha<\beta,\sigma} \left\{ U^\prime n_{i\alpha\sigma} n_{i\beta\bar{\sigma}}\right. 
 + (U^\prime-J_{\rm{H}}) n_{i\alpha\sigma} n_{i\beta\sigma}\nonumber\\
&&\left.-J_{\rm{H}}(d^\dagger_{i\alpha\sigma}d_{i\alpha\bar{\sigma}} d^\dagger_{i\beta\bar{\sigma}}d_{i\beta\sigma}
 -d^\dagger_{i\alpha\sigma}d^\dagger_{i\alpha\bar{\sigma}}
 d_{i\beta\sigma}d_{i\beta\bar{\sigma}}) \right\}.
\end{eqnarray}
where $n_{i\alpha\sigma}=d^\dagger_{i\alpha\sigma} d_{i\alpha\sigma}$.
Here,
$U$, $U^\prime$, and $J_{\rm{H}}$ respectively denote the intraorbital repulsion, the interorbital repulsion,
and the Hund's rule exchange coupling.
In the following, we will take $U^\prime=U-2J_{\rm{H}}$.~\cite{Castellani78}

\subsection{$U(1)$ slave-spin theory}
The metal-insulator transition in the model has been studied by using a $U(1)$ slave-spin theory, which was introduced in Ref.~\onlinecite{Yu-U12012}.
Here we summarize the approach to set the stage for our consideration of the OSMP in the next two sections.
For further details, we refer to Ref.~\onlinecite{Yu-U12012}
as well as Appendices~\ref{Sec:Appen_A} and \ref{Sec:Appen_B}. In addition, we refer to Appendix~\ref{Sec:Appen_C}
for a comparison with the $Z_2$ slave-spin theory of Ref.~\onlinecite{deMedici05} (see also Refs.~\onlinecite{Ruegg2010,Nandkishore2012}).

In the $U(1)$ slave-spin formulation, the XY component of a quantum $S=1/2$ spin operator ($S^+_{i\alpha\sigma}$) is used to represent
the charge degree of freedom of the electron at each site $i$, for each orbital $\alpha$
and each spin flavor $\sigma$. Correspondingly, a fermionic ``spinon'' operator
($f^\dagger_{i\alpha\sigma}$)  is used to carry the spin degree of freedom.
The electron creation operator is then represented as follows,
\begin{equation}
 \label{Eq:SSCreate} d^\dagger_{i\alpha\sigma} = S^+_{i\alpha\sigma} f^\dagger_{i\alpha\sigma}.
\end{equation}
This is implemented by a constraint,
\begin{equation}
 \label{Eq:constraint} S^z_{i\alpha\sigma} = f^\dagger_{i\alpha\sigma} f_{i\alpha\sigma} - \frac{1}{2},
\end{equation}
which restricts  the Hilbert space to the physical one.

This representation contains
a $U(1)$ gauge redundancy corresponding to
$f^\dagger_{i\alpha\sigma}\rightarrow f^\dagger_{i\alpha\sigma} e^{-i\theta_{i\alpha\sigma}}$
and $S^+_{i\alpha\sigma}\rightarrow S^+_{i\alpha\sigma} e^{i\theta_{i\alpha\sigma}}$.
Therefore, the slave spins carry the $U(1)$ charge, similarly as in the slave-rotor approach ~\cite{FlorensGeorges}.

To ensure that the saddle point captures
the correct
quasiparticle spectral weight in the non-interacting
limit (being equal to $1$), we define a dressed operator in the Schwinger
boson representation of the slave spins (in a way similar to the standard
slave-boson theory~\cite{KotliarRuckenstein}):
\begin{equation}
 \label{Eq:Zdagger} \hat{z}^\dagger_{i\alpha\sigma} = P^+_{i\alpha\sigma} a^\dagger_{i\alpha\sigma} b_{i\alpha\sigma}
 P^-_{i\alpha\sigma},
\end{equation}
where $P^\pm_{i\alpha\sigma}=1/\sqrt{1/2+\delta \pm (a^\dagger_{i\alpha\sigma} a_{i\alpha\sigma}
- b^\dagger_{i\alpha\sigma} b_{i\alpha\sigma})/2}$, and $\delta$ is an infinitesimal positive
number to regulate $P^\pm_{i\alpha\sigma}$.

Here $a_{i\alpha\sigma}$ and $b_{i\alpha\sigma}$ are Schwinger bosons representing the slave-spin operators:
$S^+_{i\alpha\sigma} = a^\dagger_{i\alpha\sigma} b_{i\alpha\sigma}$, $S^-_{i\alpha\sigma}
= b^\dagger_{i\alpha\sigma} a_{i\alpha\sigma}$, and $S^z_{i\alpha\sigma} = (a^\dagger_{i\alpha\sigma}
a_{i\alpha\sigma} - b^\dagger_{i\alpha\sigma} b_{i\alpha\sigma})/2$.
They satisfy an additional constraint,
\begin{equation}
\label{Eq:hard-core}
a^\dagger_{i\alpha\sigma} a_{i\alpha\sigma} + b^\dagger_{i\alpha\sigma} b_{i\alpha\sigma} = 1 .
\end{equation}
In other words, they are hard-core bosons.
In this representation, the constraint in Eq.~\eqref{Eq:constraint}
 becomes
\begin{equation}
\label{Eq:constraint-in-bosons}
 a^\dagger_{i\alpha\sigma} a_{i\alpha\sigma} - b^\dagger_{i\alpha\sigma} b_{i\alpha\sigma}= 2 f^\dagger_{i\alpha\sigma} f_{i\alpha\sigma} -1 .
 \end{equation}
At the same time,
Eq.~\eqref{Eq:SSCreate} becomes
\begin{equation}\label{Eq:SBcreate}
d^\dagger_{i\alpha\sigma}=\hat{z}^\dagger_{i\alpha\sigma} f^\dagger_{i\alpha\sigma}.
\end{equation}

The Hamiltonian, Eq.~ \eqref{Eq:Ham_tot},
can then be effectively rewritten as
\begin{eqnarray}
\label{Eq:HamSS} H &=& \frac{1}{2}\sum_{ij\alpha\beta\sigma} t^{\alpha\beta}_{ij}
 \hat{z}^\dagger_{i\alpha\sigma} \hat{z}_{j\beta\sigma} f^\dagger_{i\alpha\sigma} f_{j\beta\sigma}
 + \sum_{i\alpha\sigma}  (\Delta_\alpha -\mu) f^\dagger_{i\alpha\sigma}
 f_{i\alpha\sigma}
 \nonumber\\
 && - \lambda_{i\alpha\sigma}[f^\dagger_{i\alpha\sigma}
 f_{i\alpha\sigma}-\frac{1}{2}(\hat{n}^a_{i\alpha\sigma}
 - \hat{n}^b_{i\alpha\sigma})] + H^S_{\mathrm{int}}.
\end{eqnarray}
Here, $\lambda_{i\alpha\sigma}$ is a Lagrange multiplier to enforce the constraint in Eq.~\eqref{Eq:constraint-in-bosons}.
In addition,  $H^S_{\mathrm{int}}$ is the interaction Hamiltonian,  Eq.~\eqref{Eq:Ham_int},
 rewritten in the slave-spin representation
$H_{\mathrm{int}}\rightarrow H_{\mathrm{int}}(\mathbf{S})$,\cite{Yu-U12012} and
subsequently with the slave-spin operators
substituted by the Schwinger bosons.
The quasiparticle spectral weight
\begin{equation}
\label{Eq:qpWeightZ}
Z_{i\alpha\sigma}=
|z_{i\alpha\sigma}|^2 \equiv
|\langle \hat{z}_{i\alpha\sigma}\rangle|^2 .
\end{equation}
A metallic phase corresponds to $Z_{i\alpha\sigma}>0$, and a Mott insulator corresponds to $Z_{i\alpha\sigma}=0$
in all orbitals with a gapless spinon spectrum.

After decomposing the boson and spinon operators and treating the constraint on average,
we obtain two saddle-point Hamiltonians for the spinons and the Schwinger bosons, respectively:
\begin{eqnarray}
 \label{Eq:Hfmf}
 H^{\mathrm{mf}}_f &=&  \sum_{k\alpha\beta}\left[ \epsilon^{\alpha\beta}_{k} \langle \tilde{z}^\dagger_\alpha \rangle
  \langle \tilde{z}_\beta \rangle + \delta_{\alpha\beta}(\Delta_\alpha-\lambda_\alpha+\tilde{\mu}_\alpha-\mu)\right] f^\dagger_{k\alpha} f_{k\beta},\nonumber\\
 \\
 \label{Eq:HSSmf}
 H^{\mathrm{mf}}_{S} &=& \sum_{\alpha\beta} \left[Q^f_{\alpha\beta}
 \left(\langle \tilde{z}^\dagger_\alpha\rangle \tilde{z}_\beta+ \langle \tilde{z}_\beta\rangle \tilde{z}^\dagger_\alpha\right)
 + \delta_{\alpha\beta}\frac{\lambda_\alpha}{2} (\hat{n}^a_\alpha-\hat{n}^b_\alpha)\right] \nonumber\\
 &&+ H^S_{\mathrm{int}},
\end{eqnarray}
where $\delta_{\alpha\beta}$ is Kronecker's delta function,
$\epsilon^{\alpha\beta}_{k}=\frac{1}{N}\sum_{ij\sigma} t^{\alpha\beta}_{ij} e^{ik(r_i-r_j)}$, and
\begin{eqnarray}
\label{Eq:Qf}
Q^f_{\alpha\beta} &=& \sum_{k\sigma}\epsilon^{\alpha\beta}_k\langle f^\dagger_{k\alpha\sigma}
f_{k\beta\sigma}\rangle/2,\\
\label{Eq:tildez} \tilde{z}^\dagger_\alpha &=& \langle P^+_\alpha\rangle a^\dagger_\alpha b_\alpha \langle P^-_\alpha\rangle.
\end{eqnarray}
In addition,
$\tilde{\mu}_\alpha$ is an effective onsite potential defined as
\begin{equation}
\label{Eq:tilde-mu}
\tilde{\mu}_\alpha = 2\bar{\epsilon}_\alpha \eta_\alpha
\end{equation}
where
\begin{equation}
\label{Eq:tilde-bar-epsilon}
\bar{\epsilon}_\alpha = \sum_\beta\left( Q^f_{\alpha\beta} \langle\tilde{z}_\alpha^\dagger\rangle \langle \tilde{z}_\beta \rangle  + \rm{c.c.} \right)
\end{equation}
and
\begin{equation}
\label{Eq:eta}
\eta_\alpha = (2n^f_\alpha-1)/[4n^f_\alpha(1-n^f_\alpha)],
\end{equation}
with $n^f_\alpha=\frac{1}{N}\sum_k \langle f^\dagger_{k\alpha} f_{k\alpha} \rangle$.

Eqs.~\eqref{Eq:Hfmf} and \eqref{Eq:HSSmf} represent the main formulation of the $U(1)$
 slave-spin approach at the saddle-point level.
Note that the slave-spin part is single-site in nature. By contrast, the pseudofermion part must contain intersite couplings, which will play an
important role in the analysis (see next section).
We study the metal-to-insulator transitions in the paramagnetic phase preserving the translational symmetry.
These allow us to drop the spin and/or site indices
of the Schwinger bosons (slave spins) and the Lagrange multiplier $\lambda_\alpha$ in the above
saddle-point equations. We refer to
Appendices~\ref{Sec:Appen_A} and \ref{Sec:Appen_B}
for a detailed derivation of these
saddle-point Hamiltonians. The
parameters
$z_\alpha$ and $\lambda_\alpha$
are solved self-consistently.
The parameter $\tilde{\mu}_\alpha$ introduced above is crucially important to ensuring
that the noninteracting limit is properly captured (with $Z_\alpha=|\langle \tilde{z}_\alpha\rangle|^2=|z_\alpha|^2=1$
and correct electron dispersion)
regardless of whether the system is at or away from half filling
(see Appendix~\ref{Sec:Appen_C}
 for more details).
 By contrast, in the $Z_2$ slave-spin formulation,
 the parameter
 $\tilde{\mu}_\alpha$
  is absent in the saddle-point equations,
 and the proper non-interacting limit cannot be easily recovered
 for the generic case of multiple non-degenerate orbitals away from half filling;
 see Appendix~\ref{Sec:Appen_C} for further discussions on this point as well.

\section{Landau free-energy functional and the origin of the orbital-selective Mott transition}
\label{Sec:Landau}

In a multiorbital system, an OSMP may exist besides the metallic and the Mott insulating phases.
In an OSMP, some of the orbitals are Mott localized
and the others are still metallic;
the quasiparticle spectral weight $Z$ vanishes for the former orbitals and remains nonzero for the latter ones.
In this section, we clarify how an OSMP can arise in the slave-spin approach
and develop a Landau theory to describe the orbital-selective Mott transition (OSMT).

\begin{figure}[h]
\centering\includegraphics[
scale=0.3
]{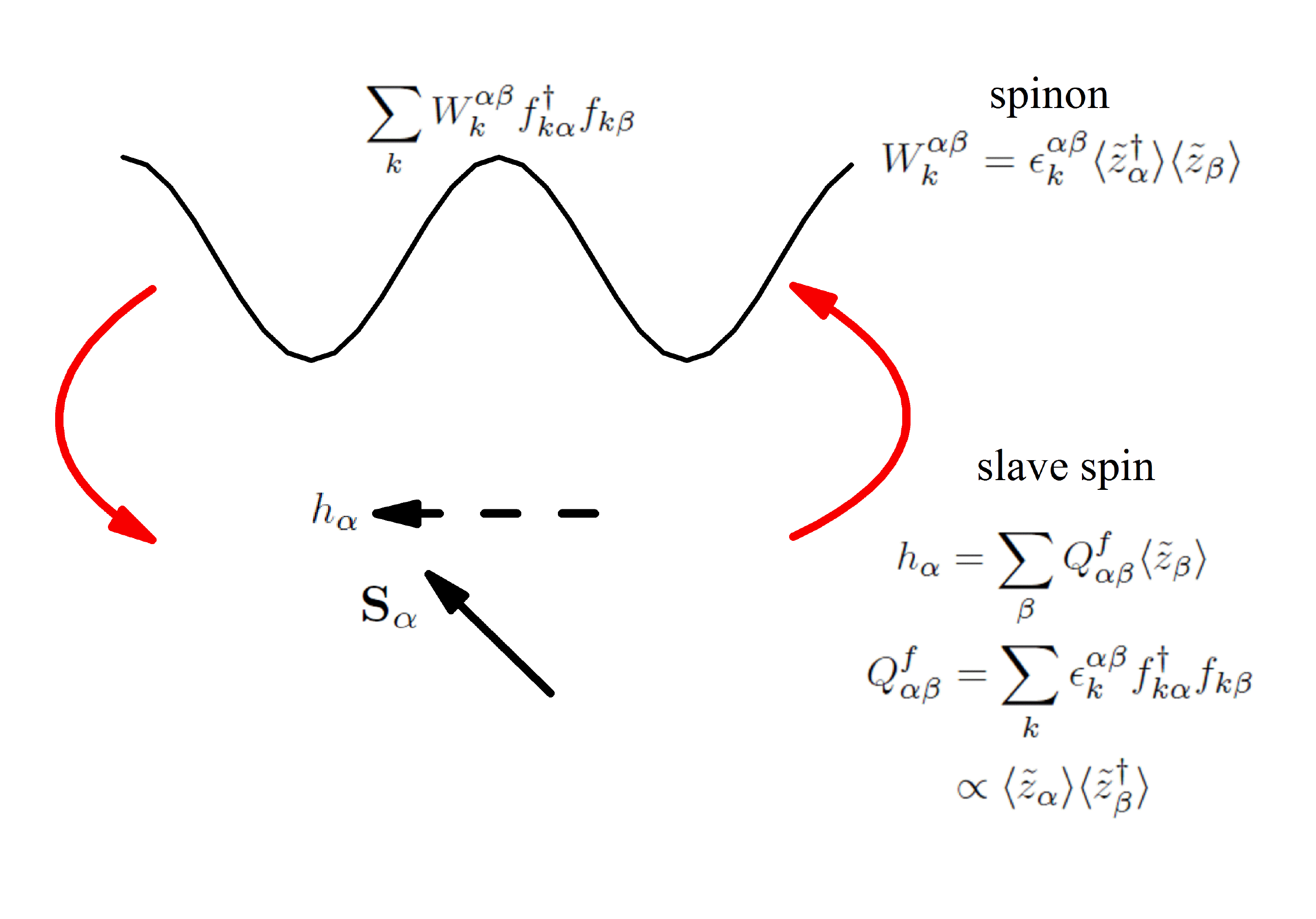}
\caption{(Color online) Illustrating
the effect of the interorbital kinetic hybridization in the $U(1)$ slave-spin theory.
The black curve shows the effective spinon dispersion,
which is generated by $ W_{k}^{\alpha\beta}f^{\dagger}_{\alpha\sigma} f_{\beta\sigma}$.
(The physical spin index $\sigma$ is suppressed in the figure legends.)
Meanwhile, the slave-spin $\mathbf{S}_\alpha$ experiences a local field,
$h_\alpha=\sum_\beta Q^f_{\alpha\beta}\langle\tilde{z}_\beta\rangle$,
where $Q^f_{\alpha\beta}\propto\langle\tilde{z}_\alpha\rangle \langle\tilde{z}^\dagger_\beta\rangle$.
The red arrows indicate
the self-consistency between $W_{k}^{\alpha\beta}$ and $h_{\alpha}$,
which results in a biquadratic interorbital coupling as shown in Eqs.~\eqref{Eq:KHf} and ~\eqref{Eq:KHS}.
}
\label{fig:1}
\end{figure}

We start from the two saddle-point Hamiltonians,
Eqs.~\eqref{Eq:Hfmf} and \eqref{Eq:HSSmf}.
Consider first Eq.~\eqref{Eq:Hfmf}, where the kinetic hybridization between two different orbitals
$\alpha \ne \beta$ is $W_{k}^{\alpha \beta} f^\dagger_{k\alpha}f_{k\beta} $,
with $W_{k}^{\alpha \beta}= \epsilon^{\alpha\beta}_{k} \langle \tilde{z}^\dagger_\alpha \rangle
\langle \tilde{z}_\beta \rangle
\propto \langle \tilde{z}^\dagger_\alpha \rangle \langle \tilde{z}_\beta \rangle$.
Recall that $\langle f^\dagger_{k\alpha}f_{k\beta}\rangle$ is determined by an averaging of
 $f^\dagger_{k\alpha}f_{k\beta}$ with respect to $H^{\rm{mf}}_f$,
 and it is nonzero in response to an effective ``field" $W_{k}^{\alpha \beta}$ applied to the kinetic
 hybridization operator $f^\dagger_{k\alpha}f_{k\beta}$ in $H^{\rm{mf}}_f$. For the case we consider,
 with a nonzero orbital level
 difference, $\Delta_{\alpha,\beta} \equiv \Delta_{\alpha}-\Delta_{\beta} \ne 0$, the
 susceptibility describing the linear response of $\langle f^\dagger_{k\alpha}f_{k\beta}\rangle$ to
 $W_{k}^{\alpha \beta}$
  will be finite, leading to
$\langle f^\dagger_{k\alpha}f_{k\beta}\rangle \propto W_{k}^{\alpha \beta}
\propto \langle \tilde{z}_\alpha \rangle \langle \tilde{z}^\dagger_\beta \rangle$;
this is illustrated in Fig.~\ref{fig:1}, top panel.
As a result, the kinetic hybridization of the spinons is
\begin{equation}
\label{Eq:KHf}
 \langle H^{\rm{mf}}_f\rangle_{\alpha\beta}=
 \sum_k
 \epsilon^{\alpha\beta}_{k}
\langle \tilde{z}^\dagger_\alpha \rangle \langle \tilde{z}_\beta \rangle \langle
f^\dagger_{k\alpha}f_{k\beta} \rangle \propto |\langle \tilde{z}_\alpha \rangle|^2 |\langle \tilde{z}_\beta \rangle|^2.
\end{equation}

Consider next Eq.~\eqref{Eq:HSSmf}, which shows that the slave-spin operator $\tilde{z}_\alpha$ for orbital $\alpha$ experiences an effective field of $h_{\alpha} = \sum_{\beta} Q^f_{\alpha\beta} \langle \tilde{z}_{\beta}\rangle$,
where $Q^f_{\alpha\beta}$ is defined in Eq.~\eqref{Eq:Qf}.
Similar reasoning as in the previous paragraph gives rise to
\begin{equation}
\label{Eq:Qf-Landau}
Q^f_{\alpha\beta}\propto \langle \tilde{z}_\alpha \rangle \langle \tilde{z}^\dagger_\beta \rangle ,
\end{equation}
for $\alpha \ne \beta$ and $\Delta_{\alpha,\beta} \ne 0$.
In other words, $h_{\alpha} \propto \langle \tilde{z}_\alpha \rangle |\langle \tilde{z}_\beta \rangle|^2$,
as illustrated in Fig.~\ref{fig:1}.
Taking the expectation value of $\tilde{z}_\alpha^\dagger$ with respect to $H^{\rm{mf}}_S$ then yields the following
component of the free energy from $H^{\rm{mf}}_S$:
\begin{equation}
\label{Eq:KHS}
\langle H^{\rm{mf}}_S \rangle_{\alpha\beta}
\rightarrow |\langle \tilde{z}_\alpha \rangle|^2 |\langle \tilde{z}_\beta \rangle|^2.
\end{equation}

From Eqs.~\eqref{Eq:KHf} and \eqref{Eq:KHS},
we see that for both the spinons and the slave spins, the interorbital correlations
appear as a biquadratic coupling $|z_\alpha |^2 |z_\beta|^2$.
This biquadratic interaction is a natural result of a self-consistent solution of the two
saddle-point equations, Eqs.~\eqref{Eq:Hfmf} and \eqref{Eq:HSSmf} (Fig.~\ref{fig:1}).
It is crucial to the stabilization of an OSMP.
As $z_\alpha $ approaches zero, so does $Q^f_{\alpha\beta}$;
correspondingly, the effective field acting on the slave spin, $h_{\alpha}$, also goes to zero
in spite of a nonzero $z_\beta$,
making the OSMP an internally consistent solution.

To see how the OSMP arises more explicitly, we can construct a Landau free-energy functional
in terms of
the quasiparticle weights,
$z_\alpha$.
 For simplicity of notation, we take the $3d$ $xy$ and another $3d$ orbital as orbitals $1$ and $2$, but our analysis straightforwardly applies to the
 case of more than two orbitals.
 The free-energy density
 reads
\begin{equation}
\label{Eq:GL} f=\sum_{\alpha=1,2} \left(r_\alpha | z_\alpha |^2 + u_\alpha | z_\alpha |^4\right) + v| z_1 |^2| z_2 |^2,
\end{equation}
in which the biquadratic coupling $v$ term comes from the kinetic hybridization as discussed above.
The quadratic terms $r_\alpha | z_\alpha |^2$ arise from the kinetic energy of the
saddle-point Hamiltonian in Eq.~\eqref{Eq:HSSmf} [as well as in Eq.~\eqref{Eq:Hfmf}].
For example, since $\sum_{k}\langle f^\dagger_{k\alpha}f_{k\alpha}\rangle$
is the spinon density in orbital $\alpha$,
which is of order $O(1)$ even when $z_\alpha $ approaches zero; thus,
$Q^f_{\alpha\alpha}\sim O(1)$,
and gives rise to the quadratic terms
($r_\alpha | z_\alpha |^2$) in the Landau free-energy density.
Taking the derivatives with respect to $|z_\alpha|$, we obtain
\begin{eqnarray}
 \frac{\partial f}{\partial |z_1|} &=& |z_1|
 (2r_1 +4u_1 |z_1|^2 + 2 v  |z_2|^2) =0,\\
 \frac{\partial f}{\partial |z_2 |} &=& |z_2|
 (2r_2 +4u_2 |z_2|^2 + 2 v |z_1|^2) =0.
\end{eqnarray}
There could then be three solutions:
\begin{enumerate}
 \item $r_1 +2u_1 |z_1|^2 + v |z_2|^2=r_2 +2u_2 |z_2|^2 + v |z_1|^2=0$, which yields $|z_1|\neq0$, $|z_2|\neq0$, corresponding to a metallic phase;
 \item $|z_1|=|z_2|=0$, corresponding to a Mott insulator;
 \item $|z_1|=0$, $|z_2|=\sqrt{-\frac{r_2}{2u_2}}$
 (or $|z_2|=0$, $|z_1|=\sqrt{-\frac{r_1}{2u_1}}$), corresponding to an OSMP.
\end{enumerate}

\section{The orbital-selective Mott transition in a five-orbital model for iron chalcogenides
as an illustration of the Landau theory}
\label{Sec:OSMP-U1SS}

In the previous section we have constructed a Landau theory, and shown that an OSMP is
an allowed solution to the free-energy functional.
Strictly speaking, the Landau theory works only when
$z_\alpha$
is sufficiently small for each orbital.
In the more realistic situation,
such as in the five-orbital model for iron chalcogenides,
$z_{xz/yz}$
may still be sizable
across the OSMT
for the localization of the $3d$ $xy$ orbital. Therefore,
it is instructive to
show  that the general
consideration
obtained in the previous section is valid in the
 five-orbital model.

\begin{figure}[h]
\centering\includegraphics[
scale=0.29
]{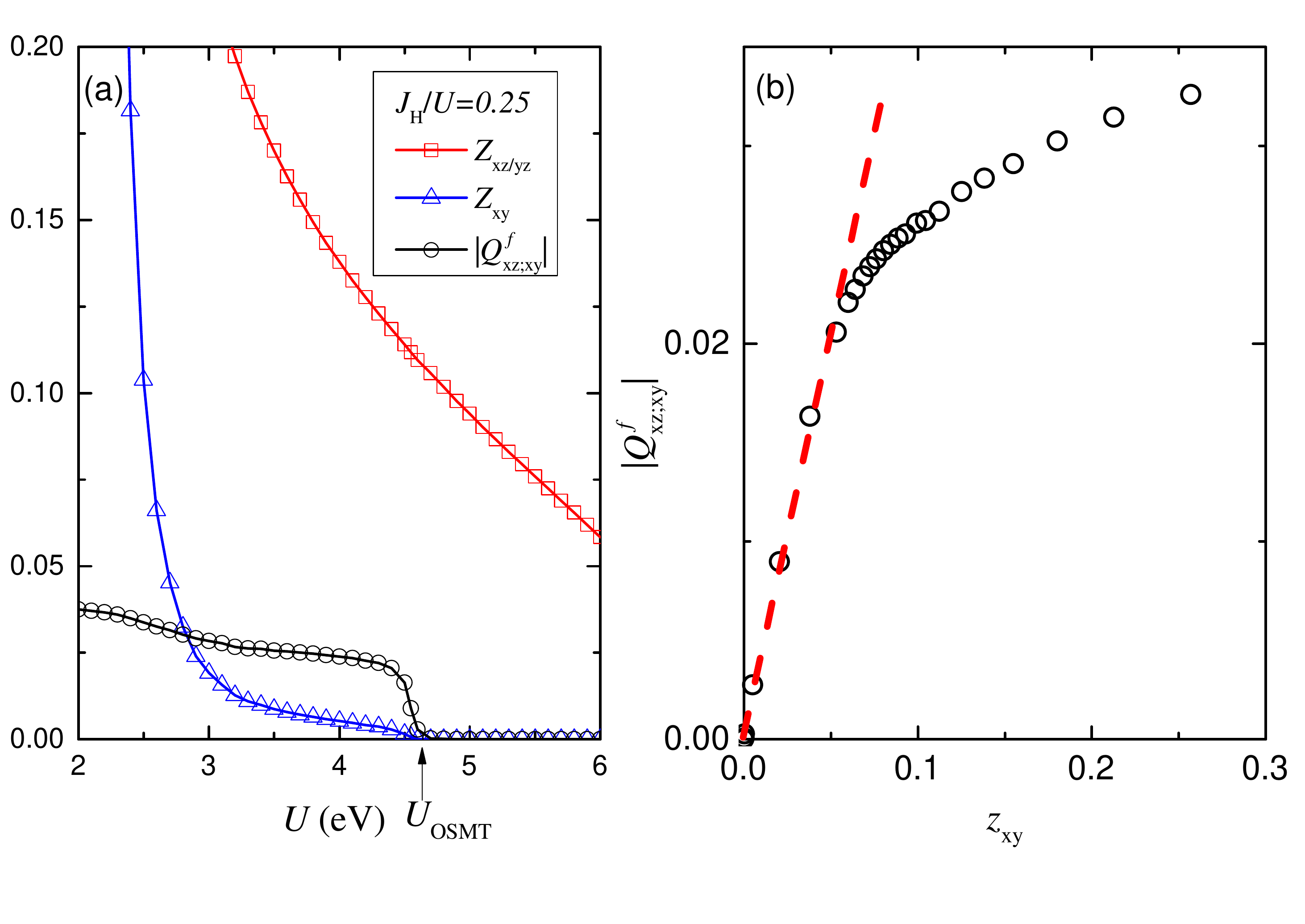}
\caption{(Color online) (a): The evolution of the
 quasiparticle spectral weights $Z_{xy}$, $Z_{xz/yz}$, and
the renormalization factor for the interorbital hybridization $|Q^f_{xz/yz,xy}|$
as a function of $U$ in a five-orbital model for K$_x$Fe$_{2-y}$Se$_2$ (without ordered iron vacancies),
with $J_{\rm{H}}/U=0.25$ and
electron filling $n=6$.
An OSMT occurs at $U=U_{\rm{OSMT}}$.
(b): $|Q^f_{xz/yz,xy}|$ vs. $z_{xy}$. The linear dependence expected from the Landau analysis
is shown by a linear fit (dashed line).
 The model parameters are the same as in (a).
}
\label{fig:2}
\end{figure}

For this reason, we revisit the OSMT in the five-orbital model for
K$_x$Fe$_{2-y}$Se$_2$ system.~\cite{Yu2013}
For an illustrative purpose, we consider the case without iron-vacancy order
and with the electron filling $n=6$.
For definiteness, we fix the ratio of the Hund's rule coupling to the intraorbital Hubbard interaction to $J_{\rm{H}}/U=0.25$.
We have
calculated the evolution of the quasiparticle spectral weights $Z_{\alpha}=|z_{\alpha}|^2$ with increasing $U$ for the
$3d$
${xz/yz}$ and ${xy}$ orbitals. The results are summarized in Fig.~\ref{fig:2}(a).
An OSMT takes place,
at which $Z_{xy}$ vanishes, where $Z_{xz}$
(as well as $Z_{yz}=Z_{xz}$, $Z_{x^2-y^2}$
and $Z_{3z^2-r^2}$)
remain
nonzero.

On approach of the OSMT,
we
can investigate  how the renormalization factor for the
interorbital hybridization, $Q^f_{xz,xy}$ (and $Q^f_{yz,xy}$, which is equal to
$Q^f_{xz,xy}$ as dictated by the $C_4$ symmetry) behaves.
The general analysis in the previous section implies that $Q^f_{xz/yz,xy}$ would vanish as the OSMT is reached.
The relationship $Q^f_{xz,xy} \propto z_{xz} z_{xy} $
would be expected
if both $z_{xy}$ and $z_{xz}$ were small.
In our case, since
$z_{xz}$ is still sizable across the OSMT,
the relationship $Q^f_{xz,xy}\propto z_{xz} $ would acquire sizable
corrections; such corrections
will not affect the existence or critical behavior of the OSMT.
On the other hand,  near the transition,
$Q^f_{xz,xy}$
is expected to be linearly
proportional to
$z_{xy}$.
This proportionality is indeed satisfied, as shown in Fig.~\ref{fig:2}(b).

\section{Discussions}
\label{Sec:Discussions}
We have analyzed the OSMT in multiorbital models pertinent to the FeSCs.
From a Landau free-energy functional,
the OSMP can be realized because the effective coupling between
the quasiparticle weights of the different orbitals is biquadratic (instead of bilinear).
In our analysis, we have emphasized the role of a nonzero orbital level splitting between
the $3d$ $xy$ orbital and the other $3d$ orbitals (particularly the $xz/yz$ orbitals).

We have clarified how the crucial feature of the Landau analysis, namely the absence
of the bilinear coupling between the quasiparticle weights of the different orbitals,
 arises within the $U(1)$ slave-spin approach. Crucial to this feature is
the coupling of the pseudofermions at different sites.
At the saddle point level, the slave spins are treated through a local description,
with its condensate capturing the quasiparticle weight. By contrast, the effective Hamiltonian of the pseudofermions,
which is at least bilinear, must contain intersite coupling.
Because the slave spins have a $U(1)$ symmetry,
we can construct the saddle-point equations in the gauge
with the slave spins alone describing the charge degrees freedom.
Correspondingly, the physical spin degrees of freedom are entirely captured by the pseudofermions.
Thus, the intersite pseudofermion coupling reflects the intersite coupling of the physical spin degrees of freedom.
In this way, the analysis here highlights the importance of the intersite spin correlations in renormalizing the interorbital kinetic hybridization to zero,
as the OSMP is realized.

This insight suggests complementary means of studying the OSMP.
In terms of physical variables,
another setting in which interorbital hybridization has been shown to be dynamically suppressed is in the heavy fermion systems \cite{Si2001,Coleman2001,Pepin2007}. There, the RKKY spin-exchange interactions compete against the Kondo hybridization
between the $f$ and conduction electrons. One of the means that captures this interplay between the hybridization
and collective spin correlations is the extended dynamical mean field theory (EDMFT) \cite{Si-jpsj2014}.
In the EDMFT approach, such intersite spin correlations are treated in terms of physical spins (instead of the auxiliary fermions).

For the multiorbital Hubbard Hamiltonian, Eq.~\eqref{Eq:Ham_tot},  one can add an explicit exchange interaction, $J_{ij}$,
between the spins at the different sites. The total Hamiltonian is then the multiorbital analogue of the one-band Hubbard-Heisenberg
model \cite{Daul-PRL2000}. (Alternatively, such a term can be considered as being effectively generated at an intermediate energy scale,
by integrating out the incoherent part of the slave-spin fluctuation spectrum.) Within the EDMFT,
the effects of such intersite correlations are studied dynamically through a bosonic bath, whose spectrum is determined self-consistently.
This coupling of the bosonic bath to the local spin degrees of freedom competes with the hybridization term.
When the competition is strong enough, it can drive the hybridization to zero, thereby realizing an OSMP.
For the Anderson lattice model relevant to heavy fermion systems, this method has already
shown that an OSMP  arises, even when
an on-site hybridization is present.
This line of study holds a clear promise to bring about further
new insights into the orbital-selective Mott physics
in multiorbital models pertinent
to the iron-based superconductors.

We close this section by noting that the orbital-selective Mott physics
is also of interest in a variety of other contexts,
such as VO$_2$ \cite{Mukherjee2016}
and multiorbital systems at low dimensions \cite{Rincon2014,Liu2016, Li2016}.

\section{Conclusion}
\label{Sec:Conclusion}

In this paper, we have revisited on the orbital-selective Mott phase identified in previous studies
in the multiorbital Hubbard
models for iron pnictides and iron chalcogenides. This type of models contains kinetic hybridization between the $3d$ electron orbitals.
We have constructed a Landau free-energy functional in terms of the quasiparticle renormalization factor, $z_{\alpha}$.
We have shown that a kinetic hybridization between the different orbitals introduces biquadratic couplings between the quasiparticle renormalization $z$ factors
of the different orbitals.
The absence of bilinear couplings
is a property that is generic to multiorbital models with nonzero orbital level splittings.
It makes the orbital-selective Mott phase possible.

Within the microscopic $U(1)$ slave-spin approach, the absence of the bilinear coupling among the $z$'s can be traced to the intersite spin correlations.
This amounts to a linear relationship between the renormalized interorbital kinetic hybridization and the quasiparticle
weight for the $3d$ $xy$ orbital (when the latter is sufficiently small).
Such a linear relationship is shown to be satisfied
in the previously identified solution near the orbital-selective Mott transition.

More generally, our analysis here illustrates that intersite spin correlations are important
in renormalizing the interorbital kinetic hybridization to zero, thereby generating the OSMP.
This insight suggests additional means of studying the orbital-selective Mott physics in the multiorbital models. In particular,
an extended dynamical mean field theory allows the study of the dynamical competition between the interorbital hybridization
and collective spin correlations. This method has been used to demonstrate
an OSMP in the case
of Anderson lattice model, which contains an on-site hybridization. It
would be a promising way to gain further insights into the dynamical suppression
 of the hybridization effect
for the orbital-selective Mott phase in multiorbital models pertinent
to the iron-based superconductors. Given the growing recognition
that the orbital selectivity plays an important role in the iron pnictides
and iron chalcogenides, such studies are clearly worth the efforts.

In short, through a Landau analysis, we have demonstrated the robustness of the orbital-selective
Mott phase in multiorbital models pertinent to the iron-based superconductors. We have also
suggested means for further theoretical studies of the orbital-selective Mott physics, which may be relevant to bad metals in a variety of correlated electron systems.

\acknowledgements
We
are grateful to E. Abrahams, L. de' Medici,  V. Dobrosavljevi\'c, Y. Komijani and G. Kotliar
for useful discussions, and to G. Kotliar for a careful reading of the manuscript and the
constructive comments to it.
This work has in part been supported by
the National Science Foundation of China Grant numbers 11374361 and 11674392
and Ministry of Science and Technology of China,
National Program on Key Research Project Grant number 2016YFA0300504 (R.Y.),
and by
the NSF Grant No.\ DMR-1611392 and the Robert A.\ Welch Foundation Grant No.\ C-1411 (Q.S.).
R.Y. acknowledges the hospitality of Rice University. Q.S. acknowledges the hospitality of University of California at Berkeley,
as well as the Aspen Center for Physics (NSF grant No. PHY-1607611)
during the 2016 summer program where part of this work was discussed.

After completing this manuscript, we became aware of another work [\onlinecite{Komijani}]
which studied a particle-hole-symmetric
multiorbital Hubbard model with an inter-orbital kinetic
and/or onsite hybridization and without an orbital-level splitting.
The two works reached consistent conclusions where there is overlap.

\appendix
\section{Derivation of the saddle-point equations}
\label{Sec:Appen_A}
To facilitate the detailed derivations and analyses presented in the next two appendices,
we summarize here the derivation \cite{Yu-U12012}
of the equations Eqs.~\eqref{Eq:Hfmf} and ~\eqref{Eq:HSSmf} from Eq.~\eqref{Eq:HamSS}.
By decomposing the slave (Schwinger) boson -- or, equivalently, the slave spin --
 and the pseudofermion operators in Eq.~\eqref{Eq:HamSS}, we obtain
\begin{eqnarray}
 \label{Eq:Hf}  H^{\mathrm{mf}}_f &=& \frac{1}{2}\sum_{ij\alpha\beta\sigma} t^{\alpha\beta}_{ij}
 \langle \hat{z}^\dagger_{i\alpha\sigma} \hat{z}_{j\beta\sigma}\rangle f^\dagger_{i\alpha\sigma} f_{j\beta\sigma}
 \nonumber\\
 && + \sum_{i\alpha\sigma}  (\Delta_\alpha - \lambda_{i\alpha\sigma}-\mu) f^\dagger_{i\alpha\sigma}
 f_{i\alpha\sigma}  ,\\
 \label{Eq:HS}  H^{\mathrm{mf}}_S &=& \frac{1}{2}\sum_{ij\alpha\beta\sigma} t^{\alpha\beta}_{ij}
 \langle f^\dagger_{i\alpha\sigma} f_{j\beta\sigma} \rangle z^\dagger_{i\alpha\sigma} z_{j\beta\sigma}\nonumber\\
 && + \sum_{i\alpha\sigma} \frac{\lambda_{i\alpha\sigma}}{2} (\hat{n}^a_{i\alpha\sigma}
 - \hat{n}^b_{i\alpha\sigma}) + H^S_{\mathrm{int}},
\end{eqnarray}
where $\langle\cdots\rangle$ denotes the averaging taken with respect to these Hamiltonians,
and
$\hat{n}^{a}_{i\alpha\sigma}=a^\dagger_{i\alpha\sigma} a_{i\alpha\sigma}$.
Here, $H^{\mathrm{mf}}_S$ has an internal $U(1)$ symmetry of the bosons.
For the single-orbital case, it is a Bose-Hubbard model for two species of hard-core bosons (or, equivalently, a model
for interacting XY spins),
and it possesses a phase
transition from a bosonic Mott insulator to a superfluid.
More generally, we start from the side with a Bose condensation
in the composite boson field $z_{i\alpha\sigma}$.
The leading term is captured by a single-site
decomposition
in Eq.~\eqref{Eq:Hf} and Eq.~\eqref{Eq:HS},
with
$\hat{z}^\dagger_{i\alpha\sigma}
\hat{z}_{j\beta\sigma} \approx \langle \hat{z}^\dagger_{i\alpha\sigma}\rangle \hat{z}_{j\beta\sigma}
+  \hat{z}^\dagger_{i\alpha\sigma} \langle \hat{z}_{j\beta\sigma}\rangle - \langle
\hat{z}^\dagger_{i\alpha\sigma}\rangle \langle \hat{z}_{j\beta\sigma}\rangle$.
We  focus on the paramagnetic phase
with the translational symmetry  preserved,
in which
 the spin and site indices
 can be dropped without causing ambiguity.
 The boson Hamiltonian then reads
\begin{eqnarray}
\label{Eq:HSsinglesite} H^{\mathrm{mf}}_S &\approx& \sum_{\alpha\beta} Q^f_{\alpha\beta}
\left(\langle \hat{z}^\dagger_{\alpha}\rangle \hat{z}_{\beta} + \langle \hat{z}_{\beta}\rangle \hat{z}^\dagger_{\alpha} \right) \nonumber\\
&& + \sum_\alpha \frac{\lambda_{\alpha}}{2} (\hat{n}^a_{\alpha} - \hat{n}^b_{\alpha}) + H^S_{\mathrm{int}}.
\end{eqnarray}
In Eq.~\eqref{Eq:HSsinglesite}, we Taylor-expand $\hat{z}_\alpha$ and $\hat{z}^\dagger_\alpha$
in terms of $\hat{A}-\langle \hat{A}\rangle$ (where $\hat{A}=\hat{n}^a, \hat{n}^b, a^\dagger b$), and keep
up to the linear terms in $\hat{A}-\langle \hat{A}\rangle$. This leads to
\begin{equation}
 \label{Eq:Zapprox} \hat{z}^\dagger_\alpha \approx \tilde{z}^\dagger_\alpha
 + \langle\tilde{z}^\dagger_\alpha\rangle \eta_\alpha [\hat{n}^a_\alpha-\hat{n}^b_\alpha-(2n^f_\alpha-1)],
\end{equation}
where $\tilde{z}^\dagger_\alpha = \langle P^+_\alpha\rangle a^\dagger_\alpha b_\alpha \langle
P^-_\alpha\rangle$. The details involved in the derivation of Eq.~\eqref{Eq:Zapprox} is given in Appendix B.
Note that $n^f_\alpha =\langle \hat{n}^a_\alpha\rangle=1-\langle \hat{n}^b_\alpha\rangle$ from the constraints. With this, we find that
$\langle \hat{z}_\alpha\rangle= \langle \tilde{z}_\alpha\rangle$, which is defined as $z_\alpha$
({\it cf.} Eq.~\eqref{Eq:qpWeightZ}).
 Using Eq.~\eqref{Eq:Zapprox}, the saddle-point Hamiltonian given in Eq.~\eqref{Eq:HSsinglesite} becomes
\begin{eqnarray}
 \label{HStilde} H^{\mathrm{mf}}_S &\approx& \sum_{\alpha\beta} Q^f_{\alpha\beta}
 \left(\langle \tilde{z}^\dagger_{\alpha}\rangle \tilde{z}_{\beta} + \langle \tilde{z}_{\beta}\rangle
 \tilde{z}^\dagger_{\alpha} \right) \nonumber\\
&& + \sum_\alpha \left(\frac{\lambda_{\alpha}}{2} + \bar{\epsilon}_\alpha
\eta_\alpha\right) (\hat{n}^a_{\alpha} - \hat{n}^b_{\alpha}) + H^S_{\mathrm{int}}.
\end{eqnarray}
Further using the constraint Eq.~\eqref{Eq:constraint}, we can redefine $\lambda_\alpha$ to move the term proportional to
$\eta_\alpha$ to $H^{\mathrm{mf}}_f$ by introducing an effective onsite potential $\tilde{\mu}_\alpha$.
We then
arrive at the two saddle-point Hamiltonians,
Eqs.~\eqref{Eq:Hfmf} and ~\eqref{Eq:HSSmf}.
Recognizing the hard-core nature of the bosons
(or, equivalently,
recognizing that they can be transformed back to XY spins),
we can exactly diagonalize
the Hamiltonian in Eq.~\eqref{Eq:HSSmf} even thought
it contains quartic terms of the boson operators in $H^S_{\mathrm{int}}$.

\section{Derivation of Eq.~\eqref{Eq:Zapprox}}
\label{Sec:Appen_B}
Starting from the definition of the projectors
$P^\pm_{i\alpha\sigma}=1/\sqrt{1/2+\delta \pm (\hat{n}^a_{i\alpha\sigma} - \hat{n}^b_{i\alpha\sigma})/2}$,
we expand $\hat{n}^a_{i\alpha\sigma} - \hat{n}^b_{i\alpha\sigma}$ about its saddle-point value and obtain
\begin{eqnarray}
\label{Eq:Ppapprox} P^+_\alpha &=&
\frac{1}{\sqrt{\left[ \frac{1}{2} +\frac{1}{2}(\langle\hat{n}^a_\alpha\rangle-\langle\hat{n}^b_\alpha\rangle)
+\delta \right]\left[ 1+\frac{\Delta\hat{n}}{1+\langle\hat{n}^a_\alpha\rangle-\langle\hat{n}^b_\alpha\rangle+2\delta} \right]}}\nonumber\\
&\approx& \frac{1}{\sqrt{\frac{1}{2} +\frac{1}{2}(\langle\hat{n}^a_\alpha\rangle-\langle\hat{n}^b_\alpha\rangle)+\delta}}
 \left\{ 1-\frac{\Delta \hat{n}}{2(1+\langle\hat{n}^a_\alpha\rangle-\langle\hat{n}^b_\alpha\rangle+2\delta)} \right\},\nonumber\\
\end{eqnarray}
where $\Delta \hat{n} = (\hat{n}^a_\alpha-\hat{n}^b_\alpha) - (\langle\hat{n}^a_\alpha\rangle-\langle\hat{n}^b_\alpha\rangle)$.
Here we have again dropped the site and spin indices for simplicity.
By using the constraint in Eq.~\eqref{Eq:constraint} we have $\langle\hat{n}^a_\alpha\rangle-\langle\hat{n}^b_\alpha\rangle=2n^f_\alpha-1$.
This further simplifies Eq.~\eqref{Eq:Ppapprox} to
\begin{equation}
 P^+_\alpha\approx \frac{1}{\sqrt{n^f_\alpha+\delta}} \left( 1-\frac{\Delta\hat{n}}{4(n^f_\alpha+\delta)} \right).
\end{equation}
Similarly, we have
\begin{equation}
 P^-_\alpha\approx \frac{1}{\sqrt{1-n^f_\alpha+\delta}} \left( 1+\frac{\Delta\hat{n}}{4(1-n^f_\alpha+\delta)} \right).
\end{equation}
Inserting these into the definition of $\hat{z}^\dagger_\alpha$ in Eq.~\eqref{Eq:Zdagger}, we obtain
\begin{widetext}
\begin{eqnarray}
 \hat{z}^\dagger_\alpha &\approx& \langle P^+_\alpha\rangle \left( 1-\frac{\Delta\hat{n}}{4(n^f_\alpha+\delta)} \right)\left[ \langle a^\dagger_\alpha b_\alpha\rangle + (a^\dagger_\alpha b_\alpha-\langle a^\dagger_\alpha b_\alpha\rangle)\right] \langle P^-_\alpha\rangle \left( 1+\frac{\Delta\hat{n}}{4(1-n^f_\alpha+\delta)} \right)\nonumber\\
 &\approx& \langle P^+_\alpha\rangle a^\dagger_\alpha b_\alpha \langle P^-_\alpha\rangle + \langle P^+_\alpha\rangle \langle a^\dagger_\alpha b_\alpha \rangle \langle P^-_\alpha\rangle \frac{(2n^f_\alpha-1)\Delta\hat{n}}{4n^f_\alpha(1-n^f_\alpha)+\delta} + O(\Delta\hat{n}^2),
\end{eqnarray}
\end{widetext}
where we have defined $\langle P^+_\alpha\rangle=1/\sqrt{n^f_\alpha+\delta}$ and $\langle P^-_\alpha\rangle=1/\sqrt{1-n^f_\alpha+\delta}$.
Using the definition of $\tilde{z}^\dagger_\alpha$ given in Eq.~\eqref{Eq:tildez}
and that of $\eta_{\alpha}$ given in Eq.~\eqref{Eq:eta}, we arrive at
\begin{eqnarray}
 \hat{z}^\dagger_\alpha &\approx& \tilde{z}^\dagger_\alpha + \langle \tilde{z}^\dagger_\alpha\rangle\eta_\alpha \Delta\hat{n} ,
\end{eqnarray}
which, then, yields
Eq.~\eqref{Eq:Zapprox}.

\section{Recovery of the proper noninteracting limit in the $U(1)$ slave-spin approach}
\label{Sec:Appen_C}
In this section we show how the saddle-point equations of the $U(1)$ slave-spin theory recovers the correct noninteracting
($U=J_{\rm{H}}=0$) limit for a general multiorbital Hubbard model. In the noninteracting limit,
the quasiparticle spectral weight of the itinerant electrons is not renormalized by the electron correlations; therefore
\begin{equation}
\label{Eq:Noninteracting_Z}
Z_\alpha=1
\end{equation}
for each orbital. In addition,
the spinon dispersion
should be identical to the original tight-binding dispersion of the physical $d$ electrons.
Both features would be captured
if Eq.~\eqref{Eq:Noninteracting_Z} is accompanied by
\begin{equation}
\label{Eq:Noninteracting_lambda}
\lambda_\alpha=\tilde{\mu}_\alpha .
\end{equation}
Indeed, in this case Eq.~\eqref{Eq:Hfmf} becomes
\begin{equation}
 \label{Eq:HfmfU0}H^{\rm{mf}}_f =\sum_{k\alpha\beta} \left\{ \epsilon^{\alpha\beta}_k
 + \delta_{\alpha\beta}(\Delta_\alpha-\mu) \right\} f^\dagger_{k\alpha}f_{k\beta}.
\end{equation}
This generates exactly the same dispersion as for the original tight-binding model of the $3d$ electrons in Eq.~\eqref{Eq:Ham_0}.

We now show that Eqs.~\eqref{Eq:Noninteracting_Z} and \eqref{Eq:Noninteracting_lambda} indeed solve the saddle-point equations
in the non-interacting case.
Our strategy is to show that $\lambda_\alpha=\tilde{\mu}_\alpha$ leads to $Z_{\alpha}=1$ in
this case.
With $\lambda_\alpha=\tilde{\mu}_\alpha$, the
Hamiltonian $H^{\rm{mf}}_S$ at $U=J_{\rm{H}}=0$ in Eq.~\eqref{Eq:HSSmf} becomes
\begin{widetext}
\begin{eqnarray}
 H^{\rm{mf}}_S &=& \sum_{\alpha\beta}\left\{ Q^f_{\alpha\beta}\left( \langle\tilde{z}^\dagger_\alpha\rangle \tilde{z}_\beta
 + \langle\tilde{z}_\beta\rangle \tilde{z}^\dagger_\alpha\right) +\delta_{\alpha\beta}
 \frac{\tilde{\mu}_\alpha}{2}(\hat{n}^a_\alpha-\hat{n}^b_\alpha)\right\}\nonumber\\
 &=& \sum_\alpha \left\{ h_\alpha\tilde{z}^\dagger_\alpha + h^*_\alpha\tilde{z}_\alpha
 + (
 h_\alpha\langle \tilde{z}^\dagger_\alpha \rangle \eta_\alpha + h^*_\alpha\langle \tilde{z}_\alpha \rangle \eta_\alpha)
 (\hat{n}^a_\alpha-\hat{n}^b_\alpha)\right\},\nonumber\\
 &=& \sum_\alpha(a^\dagger_\alpha,b^\dagger_\alpha) \mathcal{H}^{\rm{mf}}_{S\alpha} \left( \begin{matrix}
  a_\alpha\\
  b_\alpha
 \end{matrix}\right),
\end{eqnarray}
where $h_\alpha=\sum_\beta Q^f_{\alpha\beta}\langle \tilde{z}_\beta \rangle$, and $\mathcal{H}^{\rm{mf}}_{S\alpha}$ is a $2\times2$ matrix,
\begin{equation}
 \mathcal{H}^{\rm{mf}}_{S\alpha} = \left(
 \begin{matrix}
  \frac{(2n^f_\alpha-1)(h_\alpha\langle\tilde{z}^\dagger_\alpha\rangle+\rm{c.c.})}{4n^f_\alpha(1-n^f_\alpha)} &
   \frac{h_\alpha}{\sqrt{n^f_\alpha(1-n^f_\alpha)}} \\
  \frac{h^*_\alpha}{\sqrt{n^f_\alpha(1-n^f_\alpha)}} &
   \frac{-(2n^f_\alpha-1)(h_\alpha\langle\tilde{z}^\dagger_\alpha\rangle+\rm{c.c.})}{4n^f_\alpha(1-n^f_\alpha)}
 \end{matrix}\right).
\end{equation}
\end{widetext}
Without losing generality, we take $\langle\tilde{z}_\alpha\rangle$ to be real,
in which case $h_\alpha$ is also real. By diagonalizing $\mathcal{H}^{\rm{mf}}_{S\alpha}$, we obtain that, at $T=0$,
\begin{equation}
 |\langle \tilde{z}_\alpha\rangle| = \frac{1}{2\sqrt{(n^f_\alpha-1/2)^2|\langle \tilde{z}_\alpha\rangle|^2+n^f_\alpha(1-n^f_\alpha)}}.
\end{equation}
This equation has two solutions, either $|\langle \tilde{z}_\alpha\rangle|=1$ or $|\langle \tilde{z}_\alpha\rangle|=-1/(2n^f_\alpha-1)^2<0$.
Because the second solution is unphysical, we have
$|\langle \tilde{z}_\alpha\rangle|=1$,
corresponding to $Z_\alpha=1$ for each orbital $\alpha$.

Therefore,
in the noninteracting case,
$\lambda_\alpha=\tilde{\mu}_\alpha$,
and the saddle-point equations of the $U(1)$ slave-spin theory
recover the correct noninteracting limit of the multiorbital Hubbard model.
In the case of degenerate orbitals at half-filling,
this result is straightforward.
On the other hand, for the generic case of non-degenerate orbitals,
when the electron density of orbital $\alpha$ (which is proportional to $\langle\hat{n}^a_\alpha\rangle-\langle\hat{n}^b_\alpha\rangle$)
is away from half-filling, $\lambda_\alpha$ is generally nonzero,
and a nonzero $\tilde{\mu}_\alpha$ is crucial for the recovery of the proper noninteracting limit.

\subsection{Comparison with the $Z_2$ slave-spin theory}
We now make a comparison between the $U(1)$ and the $Z_2$ slave-spin theory.
In the $Z_2$ formulation, the field $\tilde{\mu}_\alpha$ is absent. Therefore, for multiorbital models with non-degenerate orbitals
away from half-filling, the saddle-point equations
of the $Z_2$ formulation do not correctly capture the limit of zero interactions.
It has been shown in Ref.~\onlinecite{Yu-U12012} that by choosing a particular projector, the saddle-point Hamiltonian $H^{\rm{mf}}_S$
of the $U(1)$ theory -- which does incorporate the parameters $\tilde{\mu}_\alpha$ --
 can be written in a form appropriate for the $Z_2$ slave-spin theory.
This suggests a route to remedy the $Z_2$ formulation in the generic case of multiorbital models with non-degenerate orbitals
away from half-filling,
such that the $\tilde{\mu}_\alpha$ parameters be introduced; whether this can be done in a natural way remains unclear.
Note that, even if this is achieved,
 the agreement between the saddle-point results of the $U(1)$ and $Z_2$ formulations
 only applies to the metallic phase where the slave spins are ordered.
The $Z_2$ theory is insufficient to describe a Mott insulating state, due to the
 fact that the pseudofermions must carry not only the physical spin degrees of freedom but also the physical charge degrees of freedom
 \cite{Nandkishore2012}.




\end{document}